\documentclass[12pt]{iopart}
\usepackage{graphicx}
\newcommand{\sign}{\mathop{\mathrm{sign}}}

\begin{document}

\title[Alpha-induced reaction cross section measurements on $^{151}$Eu]{Alpha-induced reaction cross section measurements on $^{151}$Eu for the astrophysical $\gamma$-process}

\author{Gy~Gy\"urky$^1$, Z~Elekes$^1$, J~Farkas$^1$, Zs~F\"ul\"op$^1$, Z Hal\'asz,$^1$, G~G~Kiss$^1$, E~Somorjai$^1$, T~Sz\"ucs$^1$, R~T~G\"uray$^2$, N~\"Ozkan$^2$, C~Yal\c c\i n$^2$ and T~Rauscher$^3$}

\address{$^1$ Institute of Nuclear Research (ATOMKI), H-4001 Debrecen, POB.51., Hungary}
\address{$^2$ Department of Physics, Kocaeli University, TR-41380 Umuttepe, Kocaeli, Turkey }
\address{$^3$ Department of Physics, University of Basel, CH-4056 Basel, Switzerland}

\ead{gyurky@atomki.hu}
\begin{abstract}

In order to extend the experimental database relevant for the astrophysical $\gamma$-process towards the unexplored heavier mass region, the cross sections of the $^{151}$Eu($\alpha,\gamma$)$^{155}$Tb and $^{151}$Eu($\alpha$,n)$^{154}$Tb reactions have been measured at low energies between 12 and 17\,MeV using the activation technique. The results are compared with the predictions of statistical model calculations and it is found that the calculations overestimate the cross sections by about a factor of two. A sensitivity analysis shows that this discrepancy is caused by the inadequate description of the $\alpha$+nucleus channel. A factor of two reduction of the reaction rate of $^{151}$Eu($\alpha,\gamma$)$^{155}$Tb in $\gamma$-process network calculations with respect to theoretical rates using the optical
potential by McFadden and Satchler (1966) is recommended.

\end{abstract}

\submitto{\jpg}

\section{\label{sec:intro}Introduction}

About 99\,\% of the natural nuclides beyond Iron observed in the Solar System are synthesized in processes involving neutron captures. The main component of the astrophysical s-process occurs in He-shell flashes of AGB stars \cite{booth06,gal98}. The weak s-process component takes place in pre-explosive and explosive burning in massive stars \cite{rau02}. The neutrons released in the reactions $^{13}$C($\alpha$,n)$^{16}$O and $^{22}$Ne($\alpha$,n)$^{25}$Mg are building up heavier elements through slow neutron captures and $\beta^-$-decays along the line of stability \cite{kap90,booth06}. In astrophysical environments providing very high neutron densities, neutron captures become faster than $\beta^-$-decays and synthesize extremely neutron-rich nuclides close to the neutron dripline. These nuclei decay back to stability when the neutron flux ceases \cite{arn07}. This r-process can only occur in explosive environments but the specific site or sites are yet pending identification. The details of the r-process are much less known than those of the s-process, regarding the nuclear physics input as well as the astrophysical conditions.

The remaining 1\,\% of Solar System nuclides comprises those proton rich isotopes which cannot be produced by neutron capture reactions. They are the so-called p-isotopes and 35 nuclides between $^{74}$Se and $^{196}$Hg belong to this category. In general, their production mechanism is referred to as the astrophysical p-process. Since its first consistent formulation \cite{woo78} the p-process has been thought to proceed as a $\gamma$-process, mainly by ($\gamma$,n), ($\gamma$,$\alpha$) and ($\gamma$,p) photodisintegration reactions on pre-existing s- or r-process seed isotopes in massive stars. Based solely on this model, however, the calculations are not able to reproduce the Solar System p-abundances in certain mass regions. Especially the light p-isotopes near the Mo-Ru region are largely underproduced. To alleviate this problem, several other processes have been discussed in recent years that could contribute to the production of light p-nuclei through proton captures in highly proton-rich material: the so-called rp-process \cite{sch98} and $\nu$p-process \cite{fro06}, and proton captures on highly enriched s-process seeds during a type Ia supernova explosion \cite{how91,arn03}.

The rp- (rapid proton capture) process is thought to power Galactic X-ray bursts through thermonuclear explosions on the surface of neutron stars \cite{sch98}. It may produce proton-rich nuclides up to $A\simeq 100$ \cite{sch01}. Its role in nucleosynthesis remains unclear because it is uncertain whether the produced nuclides can be expelled into the interstellar medium and play any role in the galactic chemical evolution thereafter.
Modern simulations of core-collapse supernova explosions reveal the existence of the $\nu$p-process, where in the presence of strong neutrino fluxes proton rich matter is ejected, again leading to the synthesis of nuclei through proton captures \cite{fro06}. Calculations show that this process can produce p-isotopes up to the mass region around $A=100$. Which nuclei are produced and how much mass is ejected depend sensitively on the conditions in the deep layers of a core-collapse supernova which, in turn, are closely linked to the explosion mechanism. Since the mechanism is not yet completely understood there are considerable uncertainties in the astrophysical conditions. Similar uncertainties are encountered in type Ia supernovae. In addition to the uncertainties in the explosive burning, it is also not clear yet how the required enhancement of the seed nuclei can be achieved.

All these processes may contribute to the production of the light p-isotopes, but there are many open questions regarding these processes and many details must still be worked out. On the other hand, none of these processes seem to be able to produce heavy p-isotopes for $A>100-110$. The heavy p-isotopes can only be synthesized by the classical $\gamma$-process \cite{woo78,arn03}. For the $\gamma$-induced reactions to take place, a high temperature environment is needed. It has been shown that the O/Ne layers of massive stars reach temperatures of about $2-3$ GK during the core-collapse supernova explosion for a period of the order of one second \cite{woo78,ray95,rau02}. The heavy seed isotopes are disintegrated by consecutive ($\gamma$,n) reactions complemented by ($\gamma$,$\alpha$) and ($\gamma$,p) reactions on the neutron deficient side of the nuclear chart.

Although the basics of the $\gamma$-process have been laid down several decades ago, many details of the process are still unknown and even the most recent model calculations are unable to reproduce well the Solar System abundances of heavy p-isotopes (see e.g. \cite{dil08}). On the one hand, this concerns the ambiguities in the astrophysical conditions under which the process takes place (seed isotope abundances, peak temperatures, time scale, etc.). On the other hand, large uncertainties are introduced into the calculations by the nuclear physics input, most importantly by the reaction rates (determined from cross sections). $\gamma$-process models require the use of huge reaction networks including tens of thousands of nuclear reactions, and the rates of these reactions at a given stellar temperature are necessary inputs to the network calculations. The reaction rates are generally taken from calculations using the Hauser\,--\,Feshbach statistical model \cite{rath,rtk97}. For photodisintegration reactions with charged particle emission there is only a limited number of cases in the relevant mass and energy range where the theoretical reaction rates can be compared with experimental data. Therefore the model calculations remain largely untested. Model calculations show that the resulting p-isotope abundances are very sensitive to the applied reaction rates \cite{rau06,rap06}, thus the experimental check of statistical model calculations is very important.

Experimental information about the most important $\gamma$-induced reactions can be obtained from the study of the inverse capture reactions and using the detailed balance theorem. This approach is not only technically less challenging but also provides more relevant astrophysical information than the direct study of the $\gamma$-induced reactions \cite{moh07,kiss08,rau09}. In the case of ($\alpha,\gamma$) reactions, however, the relevant energy range is well below the Coulomb barrier, making the capture cross sections very small. Therefore, in spite of the increasing experimental efforts of recent years, there are only very few nuclides for which ($\alpha,\gamma$) cross sections are available experimentally (for the list of studied isotopes see \cite{gyu10}). Moreover, these measurements are almost completely confined to the lower mass region of the p-isotopes, although calculations show that ($\gamma,\alpha$) reactions play the major role in the heavy mass region of the $\gamma$-process network \cite{rau06,rap06}. Therefore, the extension of the available experimental database of ($\alpha,\gamma$) cross sections toward the heavier mass region is of crucial importance.

In the present work the cross sections of the $^{151}$Eu($\alpha,\gamma$)$^{155}$Tb and $^{151}$Eu($\alpha$,n)$^{154}$Tb reactions have been measured. In section \ref{sec:151Eu} the importance of these reactions are underlined, section \ref{sec:experimental} gives detailed description of the experimental technique, while in section \ref{sec:results} the experimental results are presented. The results are compared with statistical model calculations and the astrophysical conclusions are drawn in section \ref{sec:discussion}.

\section{\label{sec:151Eu}The case of $^{151}$Eu + $\alpha$}

Typical $\gamma$-process model calculations \cite{dil08,rap06,rau02} show a strong underproduction of the p-isotopes at the Mo-Ru region. This underproduction may be explained (at least in part) by processes other than the $\gamma$-process (such as the above mentioned rp- and $\nu$p-processes) as they can contribute to the nucleosynthesis in this mass region. Another underproduced region, however, is in the mass range $150\leq A \leq 167$ \cite{rau02} where no alternative processes have been suggested so far. In this region no experimental ($\alpha,\gamma$) cross sections are available at all, leaving the calculated reaction rates completely untested. Several ($\alpha,\gamma$) reactions in this region show a high impact on the $\gamma$-process models \cite{rau06}, the $^{151}$Eu($\alpha,\gamma$)$^{155}$Tb reaction being one of them. Therefore, this reaction has been chosen for the present study.

\subsection{Activation method and the decay of the $^{151}$Eu\,+\,$\alpha$ reaction products}

\begin{figure}
\centering
\includegraphics[angle=-90,width=\textwidth,clip=]{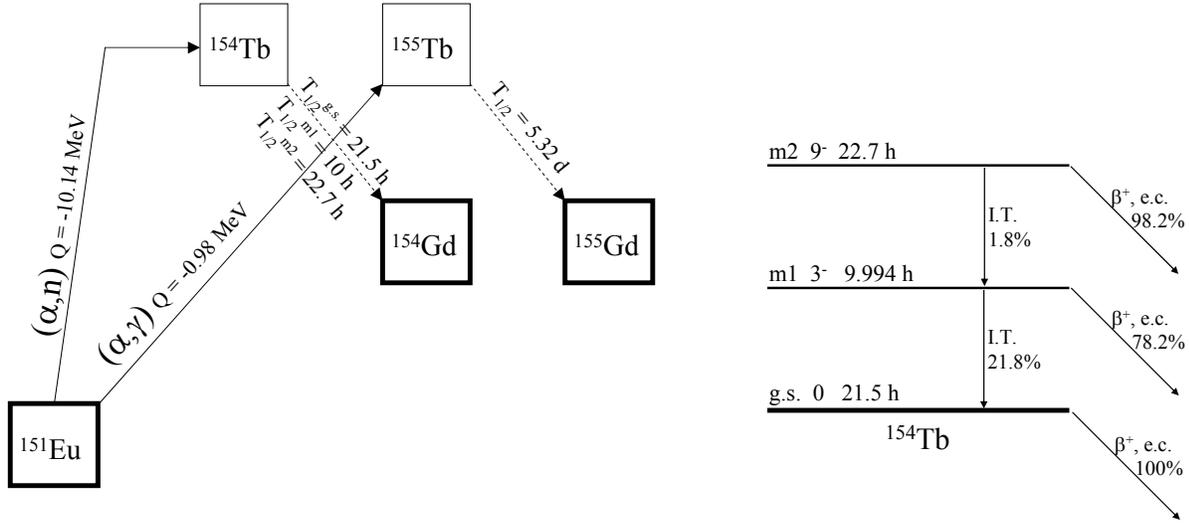}
\caption{\label{fig:decay} Left: the studied reactions and the decay of the reaction products. Right: decay of the different states in $^{154}$Tb. The decay parameters are taken from \cite{NDS05,NDS98,gyu09}.}
\end{figure}

Almost all low energy ($\alpha,\gamma$) cross section measurements for $\gamma$-process studies have been carried out using the activation technique. Although the in-beam $\gamma$-detection technique with $\gamma$-detector arrays \cite{has10} or 4$\pi$ summing crystals \cite{spy07} has also been successfully applied in some measurements, the activation method has a clear advantage in those cases where it is applicable \cite{gyu10}.

Alpha capture on $^{151}$Eu leads to $^{155}$Tb which is radioactive. It decays by electron capture to $^{155}$Gd with a half-life of 5.32\,d. The decay is followed by $\gamma$-ray emission. Therefore, the $^{151}$Eu($\alpha,\gamma$)$^{155}$Tb cross section can be measured by the activation method based on off-line $\gamma$-detection. Moreover, the $^{151}$Eu($\alpha$,n)$^{154}$Tb reaction again leads to a radioactive residual nucleus and thus the cross section of this reaction can be determined simultaneously in the activation. This reaction does not have a direct relevance for the $\gamma$-process but helps to study the optical $\alpha$+nucleus potential for the statistical model calculations (see section \ref{sec:discussion}).

The left panel of figure \ref{fig:decay} shows schematically the production and decay of the two produced isotopes. The decay of $^{154}$Tb is rather complicated and is shown in the right panel of figure \ref{fig:decay}. Besides its ground state, which decays by $\beta^+$ and electron capture to $^{154}$Gd, it has two long lived isomers. These isomers have unknown excitation energies but well established level ordering and they decay by either isomeric transition or $\beta^+$/e.c. The $^{151}$Eu($\alpha$,n) reaction can populate any of these three states. The decays of these states are followed by many $\gamma$-transitions in the $^{154}$Gd daughter nucleus. By selecting and measuring those $\gamma$-transitions which are specific for a given state only, one can measure partial cross sections leading to the various states.

Table \ref{tab:decaydata} shows some relevant decay data of the product nuclei. Only those $\gamma$-transitions are listed which have been used for the analysis. All the decay data are taken from \cite{NDS05} and \cite{NDS98} with the exception of the half-life of $^{154}$Tb$^{m1}$. The half-life of this isomer has a relatively high uncertainty in the literature: 9.4\,$\pm$\,0.4\,h, and this value is based on ambiguous data. Therefore, we have measured precisely this half-life and the updated value of 9.994\,h\,$\pm$\,0.039\,h has been used for the analysis. Details of this half-life measurement can be found in \cite{gyu09}.

\begin{table}
\caption{\label{tab:decaydata} Decay data of the reaction products. Only those $\gamma$-transitions are listed which have been used for the analysis. The decay parameters are taken from \cite{NDS05,NDS98,gyu09}.}
\begin{indented}
\item[]\begin{tabular}{lccc}
\br
Isotope & Half-life & $\gamma$-energy & Relative \\
 				&						& [keV] 					& $\gamma$-intensity [\%] \\
\mr
$^{155}$Tb & 5.23\,$\pm$\,0.06\,d & 105.3 & 25.1\,$\pm$\,1.3\\
						&											& 148.6 & 2.65\,$\pm$\,0.14\\
						&											&	180.1	& 7.45\,$\pm$\,0.41\\
$^{154}$Tb$^{g.s.}$ & 21.5\,$\pm$\,0.4\,h & 704.9 & 4.76\,$\pm$\,0.35\\
										&											& 1414.5 & 1.92\,$\pm$\,0.17\\
										&											& 2064.1 & 7.10\,$\pm$\,0.30\\
										&											& 2119.7 & 4.19\,$\pm$\,0.33\\
$^{154}$Tb$^{m1}$ & 9.994\,$\pm$\,0.039\,h& 415.9 & 2.16\,$\pm$\,0.45\\
										&											& 540.2 & 20.0\,$\pm$\,4.0\\
										&											& 1208.1 & 0.52\,$\pm$\,0.11\\
										&											& 1965.0 & 1.98\,$\pm$\,0.42\\
$^{154}$Tb$^{m2}$ & 22.7\,$\pm$\,0.5\,h		& 141.3 & 7.27\,$\pm$\,0.86\\
										&											& 225.9 & 26.8\,$\pm$\,2.8\\
\br
\end{tabular}
\end{indented}
\end{table}

\section{\label{sec:experimental}Experimental procedure}
\subsection{\label{sec:targets}Target preparation and definition}

The targets were prepared by vacuum evaporation. Eu$_2$O$_3$ powder enriched to 99.2\,\% in $^{151}$Eu has been evaporated onto 2\,$\mu$m thick Al foils. The Eu$_2$O$_3$ powder was placed into a Ta crucible heated by AC current and the Al foil fixed in a holder was placed 5\,cm above the crucible. The heated Eu$_2$O$_3$ loses oxygen and metallic Eu deposits onto the Al foil. After removal from the vacuum chamber Eu oxidizes again to Eu$_2$O$_3$. Altogether seven targets have been prepared and two different procedures have been followed. For four targets the evaporated layers were allowed to oxidize and they were used as Eu$_2$O$_3$ targets. For the other three targets a protective Al layer (typically 10\,$\mu$g/cm$^2$) has been evaporated onto the Eu layer directly after the Eu evaporation. This protective layer prevented the targets from oxidation and these targets were considered as pure metallic ones.

The number of target atoms has been determined by weighing. The weight of the Al foils was measured before and after the evaporation with a precision of about 3\,$\mu$g and from the difference the target thickness could be determined. For this method one needs the assumption that the metallic targets contain only Eu atoms and the oxide targets have the stoichiometry of Eu$_2$O$_3$. Since the oxidation of Eu takes place relatively slowly, the weight of the oxide targets was measured twice: directly after the evaporation and typically one hour later. The increase of the weight was consistent with the Eu $\rightarrow$ Eu$_2$O$_3$ transition. For testing further the target composition, the target thicknesses have also been determined by the Rutherford Backscattering (RBS) method. The targets have been bombarded with a 2.5\,MeV $\alpha$-beam from the Van de Graaff accelerator of ATOMKI and the backscattered alphas were detected by a particle detector put at 150$^\circ$ with respect to the beam direction. From the analysis of the RBS spectra the Eu target thickness was determined and found to be in agreement with the weighing results within 8\,\%. These 8\,\% have been adopted as the uncertainty of the target thickness determination. At a few energies the cross section measurements were repeated with different kind of targets (metallic or oxide) and consistent results were obtained. This also supports the reliability of the target thickness determination. The number density of $^{130}$Ba atoms were in the range between (2.5\,--\,6)$\cdot10^{17}$\,/cm$^2$. Each target was used for two or three irradiations.


\subsection{\label{sec:irradiations}Irradiations}

\begin{figure}
\centering
\includegraphics[angle=-90,width=\textwidth]{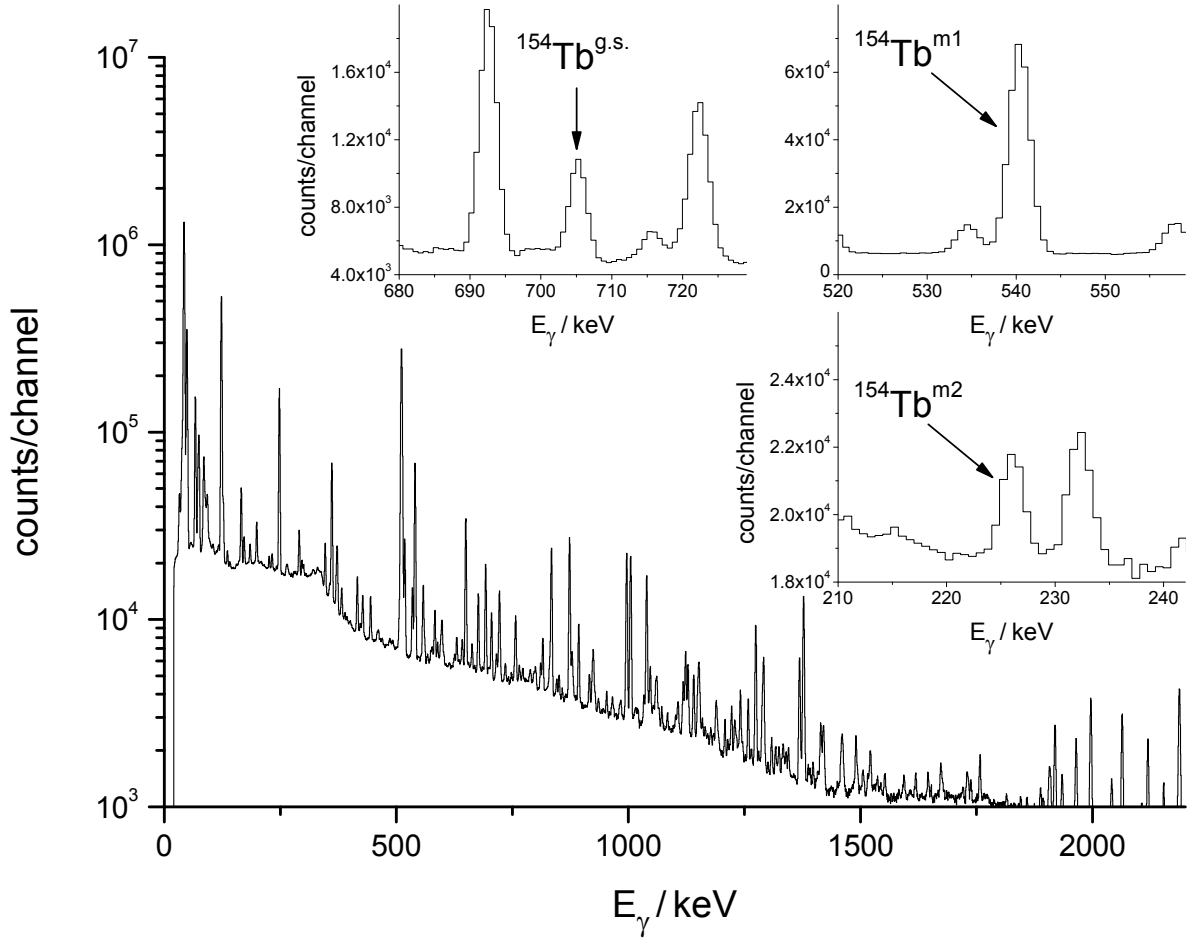}
\caption{\label{fig:shortspectrum} Gamma-spectrum measured for 12 hours after the irradiation with a 14.5\,MeV $\alpha$-beam. The very high number of $\gamma$-transitions from the decay of the three states in $^{154}$Tb is apparent. The insets show the most intense lines for the three states used in the analysis.}
\end{figure}

\begin{figure}
\centering
\includegraphics[angle=-90,width=\textwidth]{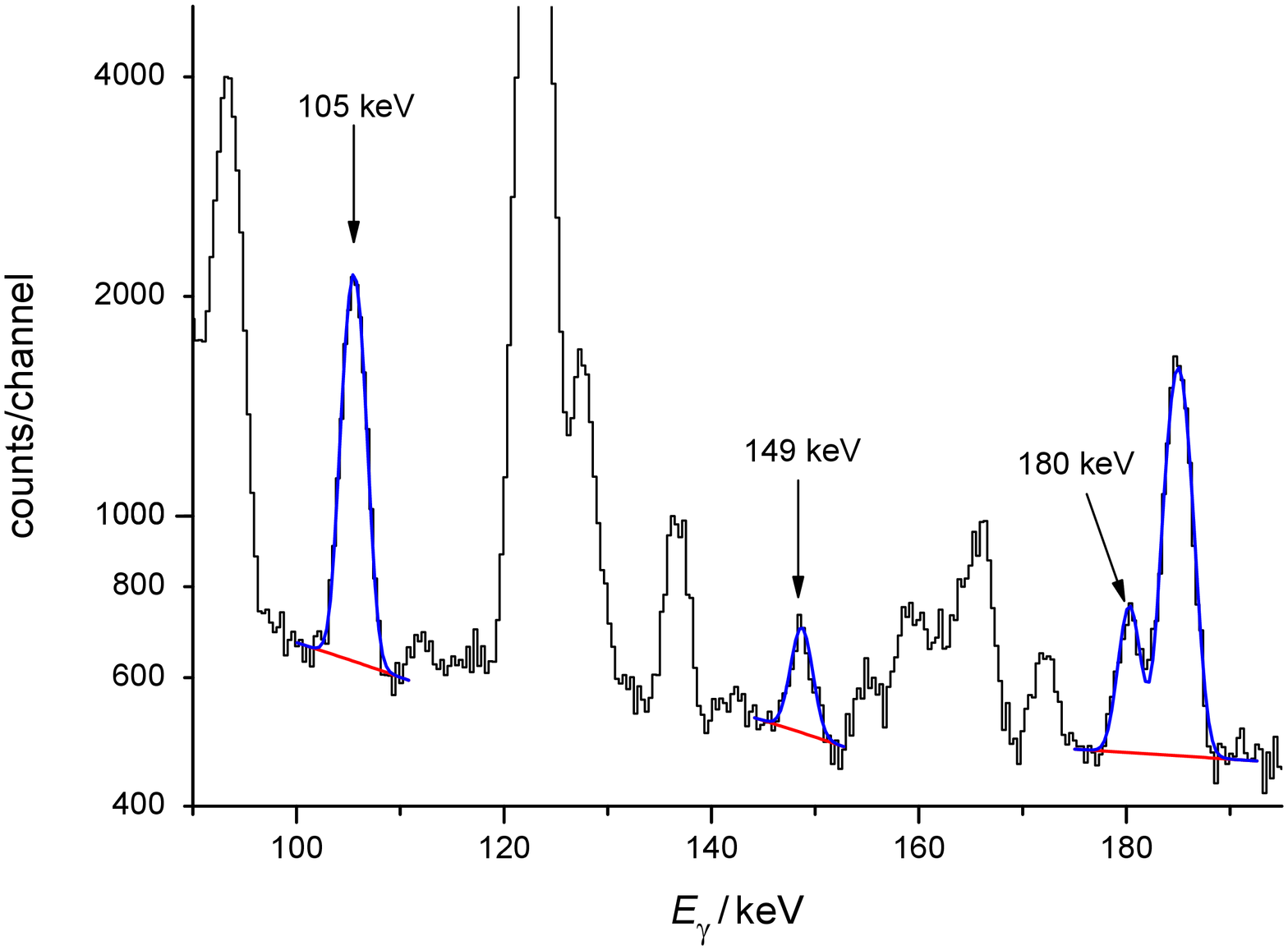}
\caption{\label{fig:longspectrum} Low energy part of the $\gamma$-spectrum measured for 18 hours on a target irradiated with a 15\,MeV $\alpha$-beam. The spectrum was taken 6 days after the irradiation when the activity of the $^{151}$Eu($\alpha$,n) reaction products were largely reduced. The lines from the $^{155}$Tb decay used in the analysis are indicated by arrows and the subtracted linear background as well as the Gaussian peak fits are also shown.}
\end{figure}

The cyclotron accelerator of ATOMKI provided the $\alpha$-beam for the activations. The target chamber is described elsewhere \cite{gyu06}. The maximum tolerable beam current on the targets has been tested using a test target with natural isotopic composition. The intensity of the 14\,MeV $\alpha$-beam was gradually increased from 0.1 to 2.5\,$\mu$A and the stability of the target was checked by detecting the backscattered $\alpha$ particles by a Si detector built into the activation chamber. No target deterioration was observed during several hours of irradiation at the maximum possible beam current. Nevertheless, RBS spectra as a function of collected charge was always measured during the activations and a beam current not exceeding 2.0\,$\mu$A was used.

The investigated energy range between E$_\alpha$\,$\approx$\,11.5 and 17.5\,MeV was covered with about 0.5\,MeV steps. The length of the irradiations varied between 5 and 24 hours. Longer irradiations were used at lower energies where the reaction cross sections are smaller. For an activation experiment the precise knowledge of the number of projectiles impinging on the target as a function of time is important. In the present work this is especially true for the $^{151}$Eu($\alpha$,n)$^{154}$Tb reaction because the half-lives of the reaction products are shorter than or comparable to the length of the irradiation. In order to follow the changes in the beam current, the digitized signals of the current integrator were recorded in multichannel scaling mode, stepping the channel in every minute. This recorded beam current profile has been used in the cross section analysis.

\subsection{\label{sec:counting}Gamma-counting}

The induced $\gamma$-activity of the targets has been measured with a 40\,\% relative efficiency HPGe detector. The countings have been started typically one hour after the end of the irradiation, lasted for 10 to 62 hours and the spectra were stored regularly (every hour) to follow the decay of the reaction products. In this period the spectrum was dominated by the decay of $^{154}$Tb and the low energy $\gamma$-lines for the longer half-life $^{155}$Tb decay were not observable because of the high Compton background. Therefore, the $\gamma$-counting of each target has been repeated typically 5\,--\,8 days after the irradiation when the $^{154}$Tb activity was already largely reduced and $^{155}$Tb could be measured. Figures \ref{fig:shortspectrum} and \ref{fig:longspectrum} show typical $\gamma$-spectra taken in the first and second counting period, respectively. In the figures the insets and the arrows show the region of the most intense $\gamma$-lines of the given isotope used in the analysis. In all cases the source activities have been determined using several $\gamma$-transitions (see table\,\ref{tab:decaydata}), consistent results were obtained and their weighted averages were adopted as the final results.

\subsection{\label{sec:efficiency}Determination of the detector efficiency}

\begin{figure}
\centering
\includegraphics[angle=-90,width=\textwidth]{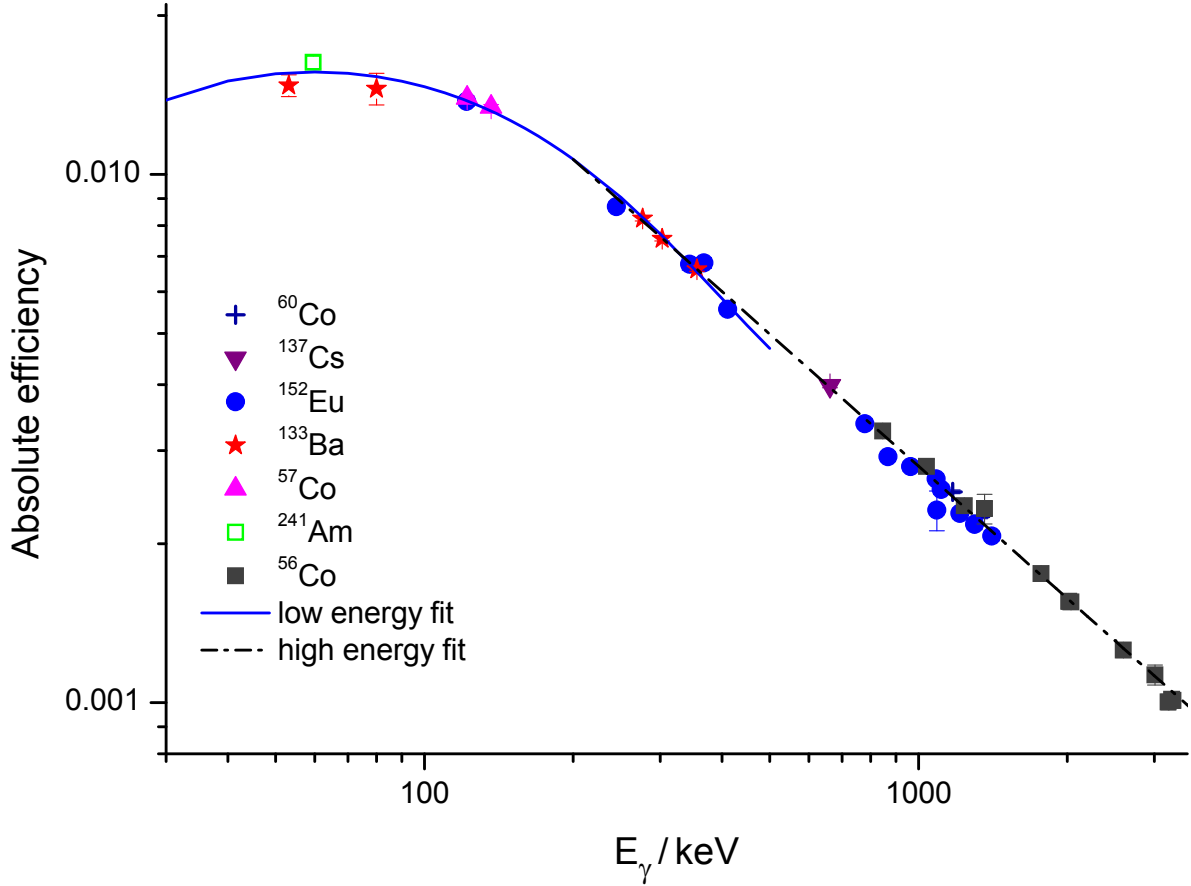}
\caption{\label{fig:efficiency} Measured and fitted absolute efficiency of the HPGe detector in far geometry. The low and high energy parts have been fitted separately.}
\end{figure}

For a precise cross section measurement the determination of the detector efficiency is of high importance. The low cross sections measured in the present work necessitated the use of a small source-detector distance and the efficiency had to be known in this geometry in a wide energy range between 105 and 2120\,keV. Moreover, the decay of all reaction products involve many cascading $\gamma$-transitions. Therefore, in close geometry the true coincidence summing effect can be significant. This holds also for any multi-line calibration sources.

This problem was circumvented by using the following procedure. First the absolute detector efficiency was measured in far geometry: at 10 cm distance from the end cap of the detector. In this geometry the true coincidence summing is negligible (below 1\,\% for the used relatively small 40\,\% detector). Six calibrated sources ($^{57}$Co, $^{60}$Co, $^{133}$Ba, $^{137}$Cs, $^{152}$Eu, $^{241}$Am) and an uncalibrated $^{56}$Co source have been used. The activity of the uncalibrated $^{56}$Co source has been determined using its $\gamma$-lines in the energy range overlapping with the other sources. Efficiency curves have been fitted to the measured points separately at the low and high energy parts. Figure \ref{fig:efficiency} shows the measured efficiencies and the fitted curves.

The actual $\gamma$-countings of the irradiated samples were carried out in a close geometry, at 3\,cm distance from the detector end cap. For all measured $\gamma$-transitions a conversion factor has been calculated between the far and close geometry efficiencies. For this purpose, strong sources of $^{154}$Tb and $^{155}$Tb have been prepared and their spectra have been measured at both close and far geometry\footnote{The $^{154}$Tb sources produced by the $^{151}$Eu($\alpha$,n)$^{154}$Tb reaction at the highest energies have been used for the $^{155}$Tb conversion factor measurement. For $^{155}$Tb, however, the sources from the $^{151}$Eu($\alpha,\gamma$)$^{155}$Tb reaction were not strong enough. Therefore, the $^{155}$Gd(p,n)$^{155}$Tb reaction has been used to produce $^{155}$Tb. This was done by irradiating evaporated natural Gd targets with 11.4\,MeV protons at the cyclotron of ATOMKI.}. Taking into account the decay between the two measurements, the conversion factor for all $\gamma$-transitions between the two geometries could be calculated. This conversion factor (its value ranges between 3 and 5 depending on the specific transition) contains both the ratio of efficiency at the two geometries and the true coincidence summing effect.

\section{\label{sec:results}Experimental results}

\begin{table}
\caption{\label{tab:agresults} Measured cross section of the $^{151}$Eu($\alpha,\gamma$)$^{155}$Tb reaction. The last three rows show the average results of the measurement carried out at the same energy.}
\begin{indented}
\item[]\begin{tabular}{lcc}
\br
E$_\alpha$ & E$^{eff}_{c.m.}$ & cross section \\
 \protect{[MeV]}                                &       [MeV]                                           & [$\mu$b] \\
\mr
12.59   &       12.25   $\pm$   0.04    &       4.93    $\pm$   1.16    \\
13.00   &       12.65   $\pm$   0.04    &       12.2    $\pm$   2.4     \\
13.50   &       13.14   $\pm$   0.04    &       18.8    $\pm$   2.7     \\
14.10   &       13.72   $\pm$   0.04    &       40.3    $\pm$   5.1     \\
14.50   &       14.11   $\pm$   0.04    &       50.7    $\pm$   6.9     \\
14.50   &       14.11   $\pm$   0.04    &       57.3    $\pm$   7.3     \\
15.09   &       14.69   $\pm$   0.05    &       99.3    $\pm$   12.7    \\
15.09   &       14.69   $\pm$   0.04    &       80.2    $\pm$   11.6    \\
15.51   &       15.10   $\pm$   0.05    &       123     $\pm$   18      \\
16.00   &       15.58   $\pm$   0.05    &       187     $\pm$   24      \\
16.00   &       15.58   $\pm$   0.05    &       190     $\pm$   27      \\
16.50   &       16.07   $\pm$   0.05    &       228     $\pm$   29      \\
17.07   &       16.62   $\pm$   0.05    &       287     $\pm$   37      \\
17.50   &       17.04   $\pm$   0.05    &       305     $\pm$   43      \\
\mr
14.50   &       14.11   $\pm$   0.04    &       52.7    $\pm$   5.9     \\
15.09   &       14.69   $\pm$   0.05    &       90.2    $\pm$   10.7    \\
16.00   &       15.58   $\pm$   0.05    &       188     $\pm$   22      \\

\br
\end{tabular}
\end{indented}
\end{table}

The astrophysically relevant energy range for the $^{151}$Eu($\alpha,\gamma$)$^{155}$Tb reaction at a typical $\gamma$-process temperature of 3\,GK is between $E_{\mathrm{c.m.}}$\,=\,7.4 and 10.4\,MeV \cite{rau10}. In this energy range, the cross section is very small and could not be measured. Therefore, the investigated energy range in the present work is a compromise between astrophysical relevance and technical feasibility. The $^{151}$Eu($\alpha,\gamma$)$^{155}$Tb excitation function has been measured from E$_\alpha$\,=\,17.5\,MeV down to the lowest possible energy of 12.5\,MeV.

The $^{151}$Eu($\alpha$,n)$^{154}$Tb reaction has a threshold at E$_\alpha$\,=\,10.4\,MeV. Above the threshold the ($\alpha$,n) channel becomes dominant quickly and its cross section becomes much higher than that of the ($\alpha,\gamma$) channel. Moreover, the decay half-life and relative $\gamma$-intensities of $^{154}$Tb are more favourable than those of $^{155}$Tb, thus the ($\alpha$,n) could be measured also at lower energies down to E$_\alpha$\,=\,11.5\,MeV, close above the threshold. Partial cross sections leading to the ground state and the two isomeric states have been determined separately. In the whole investigated energy range the population of the m1 isomer is dominant. At the lowest measured energies only upper limits could be given to the partial cross sections to the ground or m2 states.

Tables \ref{tab:agresults} and \ref{tab:anresults} show the results of the $^{151}$Eu($\alpha,\gamma$)$^{155}$Tb and $^{151}$Eu($\alpha$,n)$^{154}$Tb cross section measurements, respectively. The effective center of mass energies in the second columns take into account the energy loss of the beam in the target which was between 15 and 35\,keV. Since the cross section changes only a little in such a small energy range, the effective energy has been assigned to the center of the target thickness. The uncertainty in the center of mass energy is the quadratic sum of the beam energy uncertainty (0.3\,\%) and half of the energy loss in the target.

The uncertainty of the cross section values includes the following components: number of target atoms (8\,\%), detector efficiency at far geometry (5\,\%), conversion factor between far and close detection geometry (1\,-\,20\,\%), number of incident alphas on the target (3\,\%), decay parameters ($<$\,25\,\%), counting statistics ($<$\,20\,\%)\footnote{Higher uncertainties of the conversion factor, decay parameters, and counting statistics are typical for the weakest $\gamma$-transitions used in the analysis. The final cross sections are mainly determined by the strongest transitions for which these uncertainties are much lower.}. The cross section of a given reaction channel has been calculated based on the analysis of the different $\gamma$-transitions listed in Table \ref{tab:decaydata}. The final cross sections have been obtained by calculating the weighted mean of these data, taking into account the common and independent uncertainties. For the final uncertainties the independent uncertainties have been added quadratically. The uncertainties for each data point given in Tables \ref{tab:agresults} and \ref{tab:anresults} are not independent as they contain both the independent uncertainties (like the counting statistics) and the common systematic uncertainties.

At three bombarding energies (E$_\alpha$\,=\,14.5, 15.1 and 16.0\,MeV) two irradiations have been carried out with different type of targets (metallic or oxide). The results of these measurements, which are listed separately in the tables, are in agreement within their error bars. To give final results for these energies, the last three rows of the tables show the average cross section values of the corresponding measurements.

\begin{table}
\caption{\label{tab:anresults} Measured cross section of the $^{151}$Eu($\alpha$,n)$^{154}$Tb reaction. Partial cross sections to the ground state and two isomeric states as well as the total cross section are given. The last three rows show the average results of the measurement carried out at the same energy.}
\begin{indented}
\item[]\begin{tabular}{@{\extracolsep{-0.6mm}}lccccc}
\br
E$_\alpha$ & E$^{eff}_{c.m.}$ & \multicolumn{3}{c}{partial cross section to} & total\\
\cline{1-2}
						&									& $^{154}$Tb$^{g.s.}$ & $^{154}$Tb$^{m1}$ & $^{154}$Tb$^{m2}$ & cross section\\
 \cline{3-6}
 \multicolumn{2}{c}{[MeV]}		& \multicolumn{4}{c}{[$\mu$b]}   \\
\mr
11.65	&	11.34	$\pm$	0.04	&	$<$\,6.5			&	7.34	$\pm$	2.06	&	$<$\,3.1	&	7.3$^{+7.5}_{-2.1}$	\\
11.99	&	11.67	$\pm$	0.04	&	$<$\,8.1		&	11.6	$\pm$	2.7	&	$<$\,3.8	&	11.6$^{+9.3}_{-2.7}$	\\
12.59	&	12.25	$\pm$	0.04	&	28.9	$\pm$	5.8	&	77.7	$\pm$	15.6	&	$<$\,3.4		&	107	$\pm$	18	\\
13.00	&	12.65	$\pm$	0.04	&	62.2	$\pm$	11.9	&	211	$\pm$	42	&	$<$\,2.8	&	276	$\pm$	47	\\
13.50	&	13.14	$\pm$	0.04	&	158	$\pm$	26	&	502	$\pm$	100	&	5.97	$\pm$	2.52	&	666	$\pm$	111	\\
14.10	&	13.72	$\pm$	0.04	&	406	$\pm$	71	&	1468	$\pm$	294	&	23.9	$\pm$	7.3	&	1898	$\pm$	322	\\
14.50	&	14.11	$\pm$	0.04	&	897	$\pm$	159	&	3167	$\pm$	633	&	53.3	$\pm$	16.3	&	4117	$\pm$	698	\\
14.50	&	14.11	$\pm$	0.04	&	851	$\pm$	137	&	3329	$\pm$	666	&	55.7	$\pm$	9.8	&	4235	$\pm$	722	\\
15.09	&	14.69	$\pm$	0.05	&	2533	$\pm$	344	&	9927	$\pm$	1982	&	176	$\pm$	29	&	12636	$\pm$	2143	\\
15.09	&	14.69	$\pm$	0.04	&	3128	$\pm$	464	&	9463	$\pm$	1893	&	179	$\pm$	55	&	12770	$\pm$	2104	\\
15.51	&	15.10	$\pm$	0.05	&	3147	$\pm$	448	&	13505	$\pm$	2701	&	279	$\pm$	40	&	16932	$\pm$	2904	\\
16.00	&	15.58	$\pm$	0.05	&	8423	$\pm$	1334	&	32361	$\pm$	6473	&	719	$\pm$	217	&	41503	$\pm$	7045	\\
16.00	&	15.58	$\pm$	0.05	&	6261	$\pm$	1914	&	32415	$\pm$	6484	&	751	$\pm$	226	&	39427	$\pm$	7092	\\
16.50	&	16.07	$\pm$	0.05	&	12551	$\pm$	1779	&	55276	$\pm$	11048	&	1447	$\pm$	204	&	69274	$\pm$	11793	\\
17.07	&	16.62	$\pm$	0.05	&	13967	$\pm$	2055	&	54051	$\pm$	10811	&	1472	$\pm$	443	&	69490	$\pm$	11750	\\
17.50	&	17.04	$\pm$	0.05	&	16272	$\pm$	2570	&	80667	$\pm$	16139	&	2547	$\pm$	379	&	99486	$\pm$	17256	\\
\mr
14.50	&	14.11	$\pm$	0.04	&	870	$\pm$	120	&	3244	$\pm$	512	&	55.2	$\pm$	9.0	&	4169	$\pm$	580	\\
15.09	&	14.69	$\pm$	0.04	&	2721	$\pm$	332	&	9684	$\pm$	1527	&	177	$\pm$	27	&	12583	$\pm$	1733	\\
16.00	&	15.58	$\pm$	0.05	&	7886	$\pm$	1194	&	32388	$\pm$	5111	&	734	$\pm$	165	&	41008	$\pm$	5759	\\
\br
\end{tabular}
\end{indented}
\end{table}

\section{\label{sec:discussion}Comparison to theory}

\begin{figure}
\centering
\includegraphics[angle=-90,width=\textwidth]{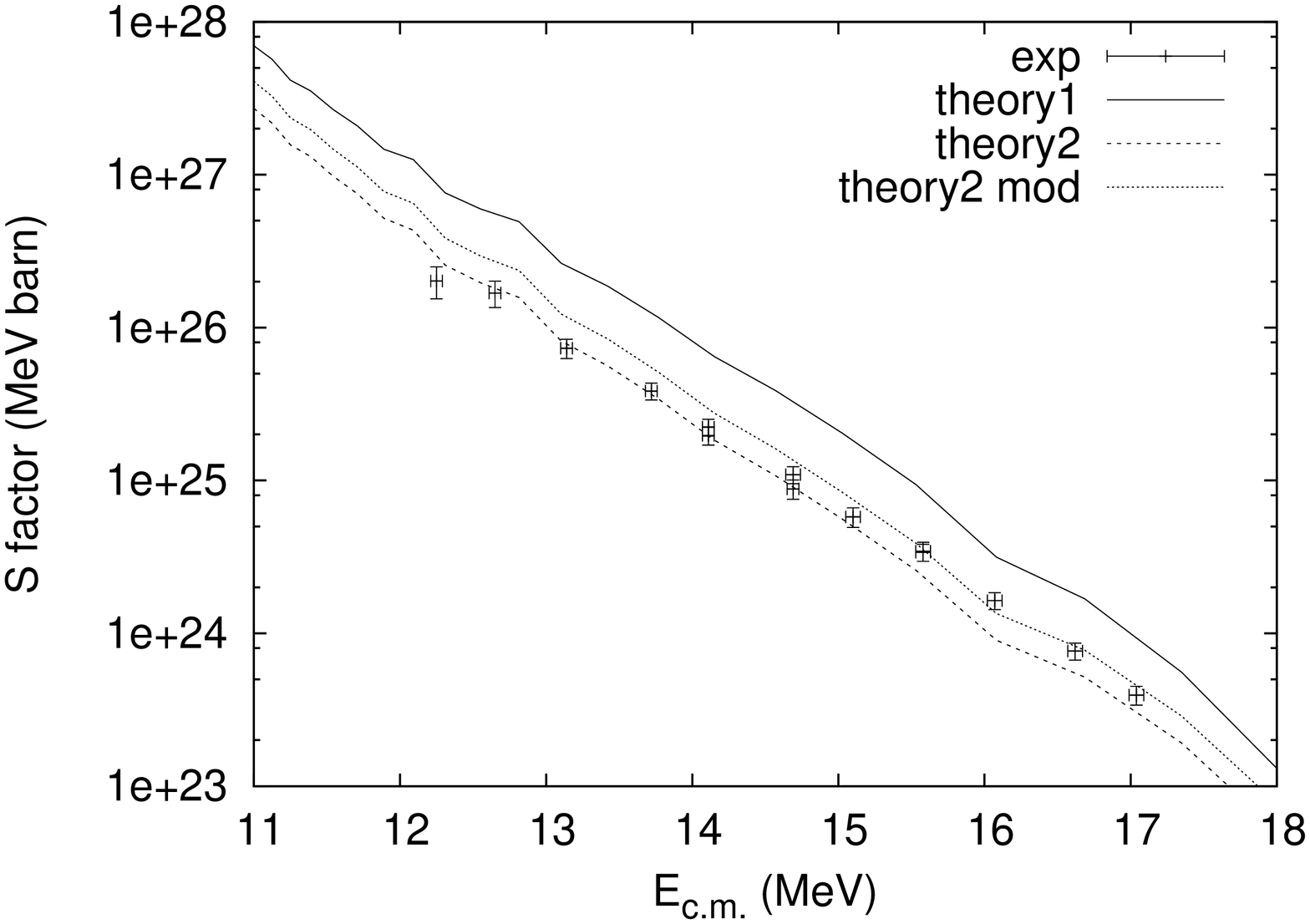}
\caption{\label{fig:ag} Comparison of experimental and theoretical astrophysical S-factors of $^{151}$Eu($\alpha$,$\gamma$)$^{155}$Tb as a function of $\alpha$ energy. The predictions were obtained with the NON-SMOKER$^\mathrm{WEB}$ code version v5.8.1, using the $\alpha$+nucleus optical potential by \cite{mcf} (theory1) and by \cite{raupot,frohdip} (theory2). The additional calculation (theory2 mod) shown was made by using the potential of \cite{raupot,frohdip} and multiplying the resulting averaged $\alpha$ transmission coefficients by a factor of 1.5.}
\end{figure}

\begin{figure}
\centering
\includegraphics[angle=-90,width=\textwidth]{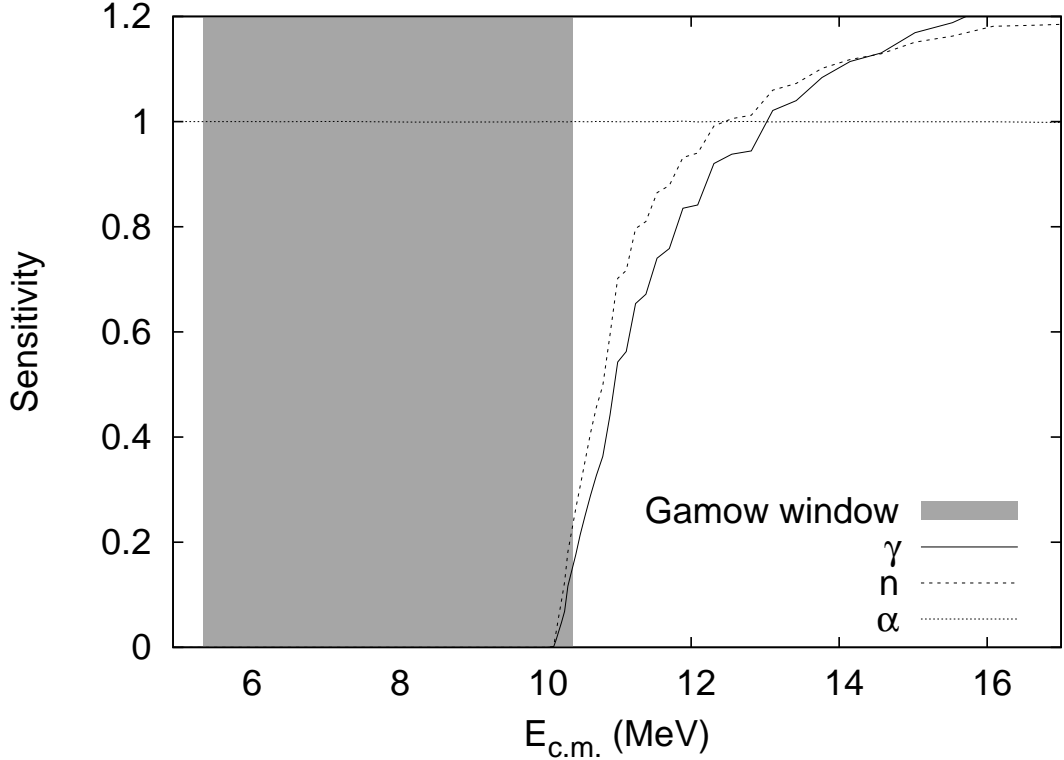}
\caption{\label{fig:sensi} Sensitivity of the $^{151}$Eu($\alpha$,$\gamma$)$^{155}$Tb cross section (and S-factor) to variations in the averaged $\gamma$-, neutron-, and $\alpha$-widths as a function of $\alpha$ energy. The sensitivity to the proton width is negligible. The shaded energy region shows the astrophysically relevant energy range (Gamow window) for the p-process for stellar temperatures $2\leq T \leq 3$ GK \cite{rau10}.}
\end{figure}

\begin{figure}
\centering
\includegraphics[angle=-90,width=\textwidth]{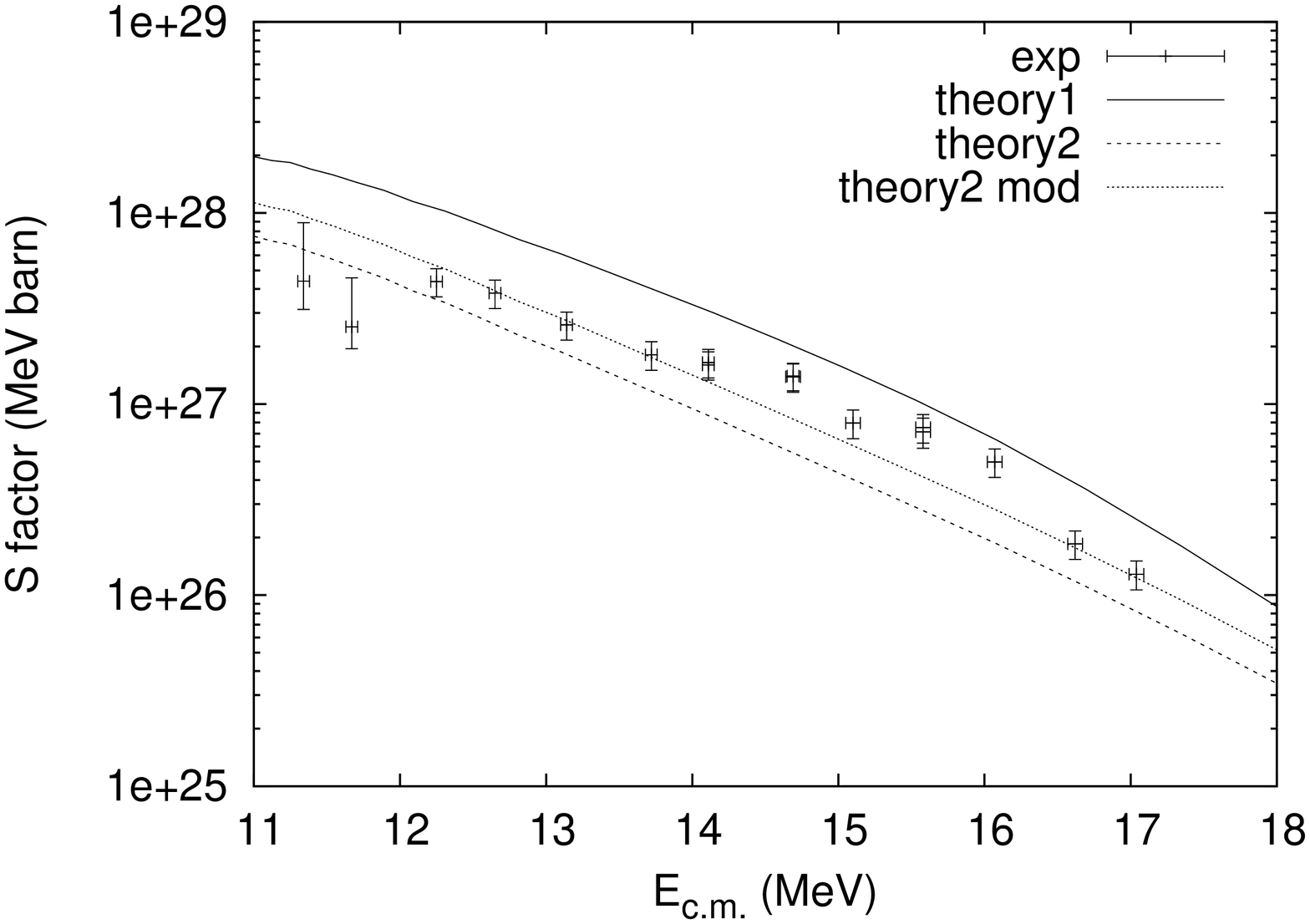}
\caption{\label{fig:an} Comparison of experimental and theoretical astrophysical S-factors of $^{151}$Eu($\alpha$,n)$^{154}$Tb as a function of $\alpha$ energy. The predictions are obtained with the NON-SMOKER$^\mathrm{WEB}$ code version v5.8.1, using the $\alpha$+nucleus optical potential by \cite{mcf} (theory1) and by \cite{raupot,frohdip} (theory2). The additional calculation (theory2 mod) shown was made by using the potential of \cite{raupot,frohdip} and multiplying the resulting averaged $\alpha$ transmission coefficients by a factor of 1.5.}
\end{figure}

As already laid out in sections \ref{sec:151Eu} and \ref{sec:results}, the reaction $^{151}$Eu($\alpha$,$\gamma$)$^{155}$Tb was chosen because it is in the mass range where ($\alpha$,$\gamma$) and ($\gamma$,$\alpha$) reactions, respectively, are important to improve our understanding of the current shortcomings in p-nuclide nucleosynthesis. The $\gamma$-process occurs in the temperature range $2\leq T\leq 3$ GK. This translates to an energy range of $5.8\leq E_\mathrm{c.m.}\leq 10.4$ MeV \cite{rau10} where cross sections are required for the calculation of astrophysical reaction rates used in reaction network calculations. Owing to the large Coulomb barrier, the cross sections become very small and currently unmeasurable in the astrophysically relevant energy range. Therefore we are forced to compare the data to predictions at higher energy. This introduces the further complication that discrepancies found between data and theory may not (or differently) appear
at astrophysically relevant energies. Therefore we have to carefully separate nuclear properties relevant in the astrophysical energy range from those not relevant.

The predictions were obtained with the statistical model code NON-SMOKER$^\mathrm{WEB}$ version v5.8.1 \cite{websmoker}. The standard prediction for the $^{151}$Eu($\alpha$,$\gamma$)$^{155}$Tb reaction compared to the data obtained in this work as a function of center of mass energy $E_\mathrm{c.m.}$ is shown in figure \ref{fig:ag} and denoted as 'theory1'. Here, we use the astrophysical S-factor $S$ which is related to the cross section $\sigma$ by
\begin{equation}
S=\sigma E e^{2\pi \eta}\quad ,
\end{equation}
with $\eta$ being the Sommerfeld parameter. The predicted S-factors are too high by a factor of about 2 across the measured energy range, with a slightly steeper energy dependence than the data.

In order to identify the cause of the discrepancy between theory and measurement it is useful to investigate the sensitivity of the S-factor to a variation of the nuclear parameters involved. The statistical model \cite{hau,gad} is making use of averaged widths $\langle \Gamma \rangle$ for the formation and decay of a compound state. The (laboratory) cross section is then given by a fraction similar as in the Breit-Wigner formula \cite{rau10,ili06}
\begin{equation}
\label{eq:fraction}
\sigma \propto \sum_i (2J_i+1) \frac{\langle \Gamma_i^\mathrm{form} \rangle \langle \Gamma_i^\mathrm{ex} \rangle}{\langle \Gamma^\mathrm{tot} \rangle} \quad .
\end{equation}
The sum runs over all compound states at the compound excitation energy $E_\mathrm{cmp}=E_\mathrm{c.m.}+E^\mathrm{sep}_\mathrm{pro}$, where $E^\mathrm{sep}_\mathrm{pro}$ is the separation energy of the projectile in the compound nucleus. The fraction includes the compound formation width $\langle \Gamma_i^\mathrm{form} \rangle$ (for the laboratory cross section this accounts for $\alpha$ transitions commencing on the ground state of the target) and the exit channel width $\langle \Gamma_i^\mathrm{ex} \rangle$. The latter sums over all energetically possible transitions to ground and excited states of the final nucleus. Finally, the total width $\langle \Gamma^\mathrm{tot} \rangle$ sums over the exit widths of all open reaction channels (including inelastic re-emission of the projectile). For the reaction $^{151}$Eu($\alpha$,$\gamma$)$^{155}$Tb, the averaged $\alpha$-width to the target ground state and the averaged $\gamma$-width in the compound nucleus $^{155}$Tb appear in the numerator of equation \ref{eq:fraction}. The denominator contains the same $\gamma$-width as the numerator but the $\alpha$-width calculated from all energetically allowed transitions to ground and excited states of the target nucleus $^{151}$Eu. Additionally, it contains the averaged proton-width and, above the threshold for neutron emission, the averaged neutron-width. Each of these widths requires different nuclear properties to be known and accordingly is sensitive to different input. For the calculation of the $\gamma$-width the excited nuclear levels and/or level density in $^{155}$Tb and the $\gamma$-strength function have to be known \cite{raugamma}. The $\alpha$-width is calculated utilizing an optical $\alpha$+$^{151}$Eu potential for transitions to known levels in $^{151}$Eu. The neutron-width requires the knowledge of the optical potential n+$^{154}$Tb and the nuclear levels in $^{154}$Tb, whereas the proton-width is calculated with the p+$^{154}$Gd potential and the levels of $^{154}$Gd. Nuclear deformation may play a role in all channels. The nuclear masses, which determine the relative energies of the contributing transitions, are known experimentally and are not free parameters.

Let us define a sensitivity $s$ as the relative change factor $f'$ of the cross section (and S-factor) when one of the widths is changed by a factor $f$, i.e.\ $\sigma'=f' \sigma$ and
\begin{equation}
\label{eq:sensi}
s=\frac{f'-1}{f-1} \quad .
\end{equation}
Therefore, $f'=(f-1)s+1$. Thus, the sensitivity tells us the impact of a change in a width on the resulting cross section. The original cross section is
$\sigma$ and the cross section obtained with a modified width is $\sigma'$. For example, if we multiply a width by a factor $f=2$ and this leads to a change in the cross section $\sigma'/\sigma=2$, then the sensitivity is $s=1$. Would the cross section be unaffected by the change in width, then the sensitivity would be $s=0$.
Here, we always assumed $\sign(f-1)=\sign(f'-1)$ but the definition can be generalized when adopting the strategy to use $1/f'$ whenever $f$ and $f'$ would indicate changes in different directions.

Figure \ref{fig:sensi} shows the sensitivities according to equation \ref{eq:sensi} for the reaction $^{151}$Eu($\alpha$,$\gamma$)$^{155}$Tb as a function of energy $Ẹ_\mathrm{c.m.}$. All widths were varied separately by a factor of two. The sensitivity to the proton-width is not plotted as it turned out to be negligible. In the measured energy range, the results are almost equally sensitive to a variation of neutron-, $\alpha$-, and $\gamma$-widths. Therefore it is impossible from this measurement alone to tell which of the width predictions is responsible for the discrepancy of the standard calculation found in figure \ref{fig:ag}. An interesting fact, however, can be seen in figure \ref{fig:sensi}: The S-factors in the astrophysically relevant energy range (as shown by the grey area) are only sensitive to the $\alpha$-width. This is due to the fact that the $\alpha$-width is strongly suppressed by the Coulomb barrier and becomes smaller than both the $\gamma$- and the proton-width towards lower energies. (The neutron-width does not contribute below the threshold.) It is then determining the S-factor because the $\gamma$-width cancels with the denominator in equation \ref{eq:fraction} (see also \cite{rau10}). Since the most relevant low-lying excited states in $^{151}$Eu are known, this implies that the knowledge of the $\alpha$+nucleus potential is essential at astrophysical energies whereas other properties are not important.

The measurement of the reaction $^{151}$Eu($\alpha$,n)$^{154}$Tb proves to be important in the above context. The S-factor of this reaction is only sensitive to the $\alpha$-width because it remains the smallest width except within a few tens of keV above the reaction threshold. A comparison of predicted and measured S-factors is shown in figure \ref{fig:an}. Again, the standard calculation (theory1) using the global $\alpha$-potential of \cite{mcf} is too high by comparable factors as for the ($\alpha$,$\gamma$) reaction. The curve labeled 'theory2' was obtained when using the potential of \cite{raupot,frohdip}. This potential was derived by fitting reaction data to three reactions at low $\alpha$ energies with compound nuclei at masses $A=144,148$. It turned out that the potential can describe low-energy reaction data across a large range of masses better then the potential of \cite{mcf} but that it fails in the reproduction of scattering data (especially at large angles) \cite{gyu06,ozk07,yal09,gal05,moh09,avri}. For the $^{151}$Eu($\alpha$,n)$^{154}$Tb case here, the use of this potential leads to a result which is closer to the data than the one obtained with the potential of \cite{mcf} but is lower than the data. Neither potential describes the features of S-factor variations seen in figure \ref{fig:an} but any statistical model calculation will give a smooth S-factor. Apart from these features, the results obtained with both potentials seem to describe the energy dependence acceptably well.

Multiplying the $\alpha$-widths obtained with the potential of \cite{raupot,frohdip} by a factor of 1.5 (curve labeled 'theory2 mod' in figure \ref{fig:an}) yields a better overall reproduction of the experimental ($\alpha$,n) data. The resulting change for the ($\alpha$,$\gamma$) reaction can be seen in figure \ref{fig:ag} where the curves for 'theory2' and 'theory2 mod' are also shown. Interestingly, there the 'theory2' results give a better overall description than the 'theory2 mod' ones. Since the 'theory2 mod' case is preferred by the ($\alpha$,n) reaction data, the remaining deficiencies have to be attributed to the treatment of the $\gamma$- and/or the neutron-width. A different treatment of these widths may result in a slightly changed slope of the S-factor function.

Since the astrophysically relevant S-factors are determined by the $\alpha$-width, we can attempt to extrapolate our findings to astrophysical energies. It has to be cautioned, however, that we thereby have to assume that the properties of the optical potential do not change at lower energy. This is not ruled out, especially the imaginary part may be strongly energy dependent with variable geometry \cite{avri,moh97,rau98,bud78}. Since there is no data to constrain the optical potential at low energy, we apply the same rescaling of the $\alpha$-widths as was found suitable to describe the ($\alpha$,n) data. This leads to the same rescaling of the S-factors and thus also of the astrophysical reaction rates. In consequence, we conclude that the $^{151}$Eu($\alpha$,$\gamma$)$^{155}$Tb rate given in \cite{rath,rath01}
(and, in fact, any prediction for this reaction using the $\alpha$ potential by \cite{mcf}) has to be reduced by a factor of two.

\section{\label{sec:summary}Summary}

We have measured the cross section of the $^{151}$Eu($\alpha,\gamma$)$^{155}$Tb and $^{151}$Eu($\alpha$,n)$^{154}$Tb reactions slightly above the relevant energy window for the astrophysical $\gamma$-process and compared the results with statistical model calculations using the NON-SMOKER$^\mathrm{WEB}$ code. With the standard parameter set the calculations for the $^{151}$Eu($\alpha,\gamma$)$^{155}$Tb reaction overestimate the measured cross sections by about a factor of two. By examining the influence of different nuclear parameters on the calculated cross section it has been found that in the astrophysically relevant energy range the calculations are mostly sensitive to the $\alpha$-nucleus optical potential, while at the measured energies other parameters play a role as well. In order to find the best $\alpha$-nucleus optical potential, the $^{151}$Eu($\alpha$,n)$^{154}$Tb reaction has been studied which even at the measured higher energies is sensitive solely to this parameter. By applying this potential to the case of $^{151}$Eu($\alpha,\gamma$)$^{155}$Tb it is suggested that the reaction rate of this reaction should be reduced by a factor of two with respect to rates from calculations using the optical potential by \cite{mcf}.

Although a new reaction rate is recommended for the $^{151}$Eu($\alpha,\gamma$)$^{155}$Tb reaction based on the present work, it is still impossible to make a general statement about how to improve the model calculations relevant for the astrophysical $\gamma$-process. This is a current challenge in nuclear astrophysics and remains unsolved for the moment. More low-energy experiments are clearly needed for both $\alpha$-capture reactions and elastic scattering especially in the mass range of heavy p-nuclei.

\ack

This work was supported by the European Research Council grant
agreement no. 203175, the Economic Competitiveness Operative Programme GVOP-3.2.1.-2004-04-0402/3.0., OTKA (K68801, NN83261), the Scientific and Technological Research Council of Turkey (TUBITAK Grant No: 108T508 and 109T585), Kocaeli University
(BAP Grant No: 2007 / 037), and the Swiss NSF (grant 2000-105328).

\end{document}